\vsize=23truecm \hsize=16.2truecm
\baselineskip=0.5truecm \parindent=0truecm
\parskip=0.2cm \hfuzz=1truecm
\font\scap=cmcsc10

\font\tenmsb=msbm10
\font\sevenmsb=msbm7
\font\fivemsb=msbm5
\newfam\msbfam
\textfont\msbfam=\tenmsb
\scriptfont\msbfam=\sevenmsb
\scriptscriptfont\msbfam=\fivemsb
\def\Bbb#1{{\fam\msbfam\relax#1}}

\newcount\eqnumber
\eqnumber=1
\def\neweq{{\rm{(\the\eqnumber)}}\global\advance\eqnumber by 1}
\def\eqdef#1{\eqno\xdef#1{\the\eqnumber}\neweq}
\def\newaeq{{\rm{(\the\eqnumber a)}}\global\advance\eqnumber by 1}
\def\eqdaf#1{\eqno\xdef#1{\the\eqnumber}\newaeq}
\def\eqdisp#1{\xdef#1{\the\eqnumber}\neweq}
\def\eqdasp#1{\xdef#1{\the\eqnumber}\newaeq}

\newcount\refnumber
\refnumber=1
\def\newref{{\the\refnumber}\global\advance\refnumber by 1}
\def\refdef#1{{\xdef#1{\the\refnumber}}\newref}

\centerline{\bf The redemption of singularity confinement}
\bigskip
\bigskip{\scap A. Ramani}
\quad{\sl Centre de Physique Th\'eorique, Ecole Polytechnique, CNRS, 91128 Palaiseau, France}
\medskip{\scap B. Grammaticos}
\quad{\sl IMNC, Universit\'e Paris VII \& XI, CNRS, UMR 8165, B\^at. 440, 91406 Orsay, France}
\medskip{\scap R. Willox} and {\scap T. Mase}\quad
{\sl Graduate School of Mathematical Sciences, the University of Tokyo, 3-8-1 Komaba, Meguro-ku, 153-8914 Tokyo, Japan }
\medskip{\scap M. Kanki}
\quad{\sl Department of Mathematics, Faculty of Science, Rikkyo University, 3-34-1 Nishi-Ikebukuro, 171-8501 Tokyo, Japan}
\bigskip
{\sl Abstract}
\smallskip
We present a novel way to apply the singularity confinement property as a discrete integrability criterion. We shall use what we call a full deautonomisation approach, which consists in treating the free parameters in the mapping as functions of the independent variable, applied to a mapping complemented with terms that are absent in the original mapping but which do not change the singularity structure. We shall show, on a host of examples including the well-known mapping of Hietarinta-Viallet, that our approach offers a way to compute the algebraic entropy for these mappings exactly, thereby allowing one to distinguish between the integrable and non-integrable cases even when both have confined singularities.
\bigskip
Singularity confinement [\refdef\sincon] was proposed nearly twenty-five years ago as a criterion for detecting integrability in discrete systems. The simple idea behind the singularity confinement requirement was that singularities ought to play an important role in the integrability of discrete rational systems, just as they do in the continuous case. In order to fix the ideas concerning the terms ``singularity'' and ``confinement'', it is best to start from a simple example. Let us therefore  first consider the mapping
$$x_{n+1}+x_{n-1}=x_n+{1\over x_n}.\eqdef\eqi$$
Clearly, when $x_m$, at some iteration $m$, takes the value 0 something special happens to the mapping. Iterating further, we find the succession of values $x_{m+1}=\infty$ and $x_{m+2}=\infty$, after which $x$ becomes indeterminate since we encounter an expression of the type $\infty-\infty$. We therefore consider the value $x_m=0$ here as a singularity of the mapping (\eqi). In general, we are in the presence of a singularity whenever $x_m$ is such that the value of $x_{m+1}$ does not depend on the value of $x_{m-1}$. Sometimes we refer to this situation as ``a loss of a degree of freedom'' for the mapping. The notion of confinement is associated with the removal of the indeterminacy that arises because of a singularity. The way to do this is through an argument of continuity with respect to the initial conditions. Instead of assuming that $x_m=0$, we introduce a small quantity $\epsilon$ and iterate the mapping starting from $x_m=\epsilon$. By taking the limit $\epsilon\to0$ we recover for $x_m,x_{m+1},x_{m+2}$ the values obtained above but it turns out that the value of $x_{m+3}$ is no longer indeterminate but equal to 0. Iterating further we find that $x_{m+4}$ is well defined and, in fact, equal to $-x_{m-1}$. Since the value of $x_{m-1}$ has reappeared through the limiting procedure, we claim that the mapping has recovered its lost degree of freedom. We refer to this removal of indeterminacy with recovery of the degree of freedom as the {\it confinement of the singularity}.

Singularity confinement was instrumental in transforming the entire domain of discrete integrable systems in that it proved to be of invaluable importance in the derivation of a multitude of new integrable examples. In particular, it enabled the derivation of discrete analogues of the Painlev\'e equations through what we called the deautonomisation procedure [\refdef\desoto]. This procedure consists in obtaining non-autonomous extensions of integrable mappings, by using some suitable integrability criterion. 

Despite these successes, the usefulness of singularity confinement was put in doubt when an example of a non-integrable mapping with confined singularities was announced. As shown by Hietarinta and Viallet [\refdef\hiv], the mapping (which we refer to as the H-V mapping)
$$x_{n+1}+x_{n-1}=x_n+{1\over x_n^2}\eqdef\eqii$$
has exactly the same pattern of singularities as (\eqi), namely $\{0, \infty, \infty, 0\}$, and its singularity is confined with $x_{m+4}=x_{m-1}$. However, (\eqii) is not integrable and this led the authors of [\hiv] to propose another criterion for discrete integrability. Based on the observation that the growth properties of the solution of a mapping are related to the integrable character of the latter, Hietarinta and Viallet introduced the notion of algebraic entropy: if $d_n$ represents the homogeneous degree of the numerator or denominator of $x_n$, the algebraic entropy of the mapping is given by the limit $\varepsilon=\lim_{n\to\infty}{1\over n}\log d_n$. While for an integrable mapping the algebraic entropy must vanish, a non-zero value for $\varepsilon$ is considered to be an indication of non-integrability. For example, in the case of mapping (\eqii) we find $\varepsilon=\log({3+\sqrt 5\over 2})$. 

In this paper we shall reconsider the trustworthiness of singularity confinement as an integrability criterion. Based on recent findings which link the behaviour of the solutions of a mapping to that of its coefficients  [\refdef\mase], we shall show that when one considers the non-autonomous extensions of mappings with confined singularities, one obtains a clear indication of the integrability or non-integrability of the mapping at hand and, in fact, that one can even compute its algebraic entropy exactly. 

In [\mase] various non-autonomous mappings obtained from so-called late confinement -- in which the singularities are not required to be confined at the earliest possible stage but only at some subsequent opportunity -- were studied using algebro-geometric techniques. In particular, it was shown that for each of these mappings, the constraints on their parameters, as obtained from singularity confinement, were in fact equivalent to the linear transformation induced, by the mapping, on part of the Picard group of the (family of) rational surfaces on which it can be regularised by blowing-up from $\Bbb{P}^1\times\Bbb{P}^1$. As the same phenomenon was observed for every single mapping we studied, be it an integrable or a non-integrable one, we conjectured that for any confining mapping of the plane, the behaviour of the solutions of the confinement constraints (and thus of the parameters in the mapping) will be governed by the linear action on the Picard group that is obtained after successfully blowing-up the mapping. The fact that knowledge of the action on the Picard group allows one to calculate the algebraic entropy of the mapping rigorously [\refdef\take]  then naturally leads to the conjecture that if the deautonomisation of a confining mapping is sufficiently general so  that its parameters will depend on the largest eigenvalue of the linear map on the Picard group, the value of this eigenvalue and hence also the algebraic entropy of the mapping can be, so to speak, read-off directly from the confinement constraints themselves. 

We therefore conjecture that singularity confinement applied in conjunction with what we call a full deautonomisation (a term to be defined later in this article) is in fact a sufficient criterion for discrete integrability. We shall give several examples of this approach, concentrating on mappings that are not of QRT type, as these fall outside the class of mappings studied in [\mase]. 

\medskip
{\sl Two confining, non-integrable, mappings}
\smallskip
In [\refdef\mimura] we introduced the mapping
$$x_{n+1}x_{n-1}={x_n^4-1\over x_n^4+1},\eqdef\eqiii$$
which has confined singularities although it is non-integrable. Its singularity patterns 	are $\{\pm1,0,\mp1\}$, $\{\pm i,0,\pm i\}$, $\{\pm r,\infty,\mp ir\}$ and $\{\pm ir,\infty,\mp r\}$ where $r$ is the square root of $i$, i.e. $r^2=i$. The non-integrability of this mapping can be assessed by means of the algebraic entropy criterion. By computing successive iterates of the mapping and obtaining their homogeneous degrees, we find empirically that the algebraic entropy of (\eqiii) is equal to $\varepsilon=\log(2+\sqrt3)$. Another piece of evidence for the non-integrability of (\eqiii) is furnished by the Nevanlinna theory for discrete systems. As shown in [\refdef\nevan], a mapping of the form of (\eqiii) cannot be integrable if the maximal degree of the numerator and denominator of the right-hand side exceeds 2. 

How can this non-integrable character be reconciled with the confinement property? The answer is to be found in the deautonomisation approach. We start by extending (\eqiii), minimally, by  introducing a function in the numerator:
$$x_{n+1}x_{n-1}={x_n^4-q_n^4\over x_n^4+1}.\eqdef\eqiv$$
This function is to be determined by requiring that the singularity patterns remain the same as for the autonomous case. We obtain readily the confinement condition
$$q_{n+1}q_{n-1}=q_n^4.\eqdef\eqv$$
Putting $\log q_n=\lambda^n$ we find for (\eqv) the characteristic equation $\lambda^2-4\lambda+1=0$ and the growth of the function $q_n$ is such that $\lim_{n\to\infty}{1\over n}\log\log q_n=\log(2+\sqrt3)$. This is precisely the value of the algebraic entropy for (\eqiii). Thus using the conjectured relation between the growth of the solutions of the mapping and that of the coefficients, we can obtain its algebraic entropy exactly and conclude on the non-integrable character of (\eqiii).

Another interesting example is the mapping introduced in [\refdef\tsuda]:
$$x_{n+1}=x_{n-1}\left(x_n-{1\over x_n}\right).\eqdef\eqvi$$
This mapping again has confined singularities, the pattern being $\{\pm1,0,\infty,\mp1\}$. Its dynamics are, however, non-integrable and its algebraic entropy is equal to $\varepsilon=\log({1+\sqrt 5\over 2})$. Note that in this case the Nevanlinna approach does not allow one to reach a similar conclusion since the degrees of the right-hand side are quite low.

Just as in the previous case we extend the mapping by introducing a function in the numerator of the right-hand side:
$$x_{n+1}=x_{n-1}\left(x_n-{q_n^2\over x_n}\right).\eqdef\eqvii$$
Again we determine this function by requiring the same confinement as in the autonomous case, which results in the confinement condition
$$q_{n+2}=q_n^2q_{n-1}.\eqdef\eqviii$$
Seeking a solution of the form $\log q_n=\lambda^n$, we find the characteristic equation: $(\lambda^2-\lambda-1)(\lambda+1)=0$. The growth of $\log q_n$, with $n$, is therefore again exponential and we find $\lim_{n\to\infty}{1\over n}\log\log q_n=\log({1+\sqrt 5\over 2})$, i.e. precisely the value of the algebraic entropy.

The conclusion we can draw from these two examples is particularly simple. Given a mapping with confined singularities and for which there exists a doubt concerning its integrable character, one should try to deautonomise the mapping. If by deautonomising we find a characteristic polynomial with some root with modulus larger than 1, we should take this as an indication that the mapping is non-integrable. In this case, as a bonus, the largest root (in absolute value) will furnish the algebraic entropy of the mapping. On the other hand, should the mapping turn out to be integrable after all, the deautonomisation used in this analysis will of course constitute an interesting extension of the original mapping.
\medskip
{\sl The H-V mapping, revisited}
\smallskip
Next we turn to the H-V mapping, since it constitutes the best known example of a non-integrable confining system. Following the approach introduced above we seek to deautonomise the mapping by introducing a function (that remains to be determined) in the right-hand side:
$$x_{n+1}+x_{n-1}=x_n+{q _n\over x_n^2}.\eqdef\eqix$$
By applying singularity confinement we obtain the known result  [\take] that $q_n$ is a purely periodic function with period 3. However, as can be understood from the analysis given in [\take], this is not the only extension of (\eqii) compatible with the singularity pattern $\{0,\infty,\infty,0\}$. In fact, adding a term inversely proportional to $x$ and/or a term independent of $x$ to the right-hand side does not modify the confinement pattern. It turns out that a term independent of $x$  can be added provided it is proportional to $(-1)^n$. The case of a term inversely proportional to $x$,
$$x_{n+1}+x_{n-1}=x_n+{f_n\over x_n}+{1\over x_n^2},\eqdef\eqx$$
is more interesting however.
Requiring the singularity pattern to remain the same, we obtain for $f_n$ the constraint:
$$f_{n+3}-2f_{n+2}-2f_{n+1}+f_{n}=0.\eqdef\eqxi$$
Taking $f_n=\lambda^n$ we find the characteristic equation $(\lambda^2-3\lambda+1)(\lambda+1)=0$, the largest root of which is ${3+\sqrt 5\over 2}$. The logarithm of this root is precisely the value of the algebraic entropy for the H-V mapping. Thus while $f_n=0$ is a possible solution of (\eqxi), just as $q_n=1$ was a possible solution of (\eqv) and (\eqviii), in order to obtain the most general deautonomisation, the full solution of (\eqxi) must be considered and then we have a clear indication of the non-integrability of the mapping as we obtain an estimate of its algebraic entropy (an estimate, which here is indeed the precise value). 

Compared to the cases (\eqiii) and (\eqvi), here, the deautonomisation is carried one step further. Namely, we consider terms that are initially {\sl absent} from the equation but the presence of which would not alter the singularity pattern. Essentially, what we do is replace a zero coefficient for such terms with a function that is to be determined by the confinement constraints. It turns out, in this precise example, that there exists a term that, when present, portends the non-existence of integrability. What we claim is that this is not a mere coincidence, specific to the case of (\eqix), but something that is widely applicable. We shall call this process of deautonomising, while adding terms that do not modify the initial singularity pattern, ``full deautonomisation''. 
\medskip
{\sl Extending the H-V mapping}
\smallskip
In order to provide further arguments in support of this claim we consider the following extension [\refdef\kanki] of the H-V mapping
$$x_{n+1}+x_{n-1}=x_n+{q_n\over x_n^k},\eqdef\eqxii$$
where $k>2$ and where $q_n$ is a function to be determined so that the singularity pattern of (\eqxii) is the same as that of the initial H-V mapping, i.e. $\{0,\infty,\infty,0\}$. The condition we obtain from this requirement  is 
$$q_{n+3}-(-1)^kq_n=0.\eqdef\eqxiii$$
In the case of an even $k$ the simple solution $q_n=1$ is sufficient for the singularity to confine. For odd $k$ we can simply take $q_n=(-1)^n$. In both cases the full solution of (\eqxiii) also has a period-3 component, just as in the case of (\eqix), but since it will not play any role in what follows we may as well neglect it.

Given the form of the right-hand side of (\eqxii) it is clear that we can add terms proportional to $x_n^{-\ell}$, with $\ell=0,\cdots,k-1$, while preserving the singularity pattern of the initial system. We are not going to go here through a detailed presentation of all the possibilities. It turns out that the $1/x_n$ term, which was the tip-off in the case of (\eqix), will again provide a telltale sign of the non-integrability of the mapping. We start from 
$$x_{n+1}+x_{n-1}=x_n+{f_n\over x_n}+{q_n\over x_n^k},\eqdef\eqxiv$$
with $q_n$ being chosen accordingly, depending on the parity of $k$, and where $f_n$ is the function that needs to be determined.
The confinement condition turns out to be
$$f_{n+3}-kf_{n+2}-kf_{n+1}+f_{n}=0,\eqdef\eqxv$$
and the ansatz $f_n=\lambda^n$ leads to the characteristic equation $(\lambda^2-(k+1)\lambda+1)(\lambda+1)=0$. The largest root of the latter is ${1\over2}(k+1+\sqrt{(k-1)(k+3)})$, in perfect agreement with the results on algebraic entropy in [\kanki] for even $k$.
\medskip
{\sl A case of late confinement}
\smallskip
Another instance where non-integrable mappings with confined singularities appear is that of ``late'' confinement [\refdef\late]. The standard practice when deautonomising a mapping with the help of the singularity confinement criterion, is to enforce the confinement contraints at the very first opportunity. It is however possible to ignore this first confinement opportunity and to try to confine at a later stage. This leads invariably to a non-integrable system. In [\mase] we have presented the algebro-geometric justification for this phenomenon. Here we shall illustrate this through a detailed example, which will further support the present singularity confinement based approach. We start with the mapping
$$x_{n+1}+x_{n-1}={f_n\over x_n}+{1\over x_n^2}.\eqdef\eqxvi$$
The ``standard'' singularity pattern of (\eqxvi) is $\{0,\infty,0\}$. If we require confinement with the standard singularity pattern we obtain the constraint $f_{n+1}-2f_n+f_{n-1}=0$, the solution of which is $f_n=\alpha n+\beta$ leading to a discrete analogue of the Painlev\'e I equation.

However, if we ignore this first opportunity to confine the singularity, another opportunity appears after four more iterations (and, in fact, infinitely many opportunities appear after adding any multiple of four steps). The confined singularity pattern now becomes $\{0,\infty,0,\infty,0,\infty,0\}$ and the confinement constraint is
$$f_{n+5}-2f_{n+4}+f_{n+3}-f_{n+2}+f_{n+1}-2f_n+f_{n-1}=0.\eqdef\eqxvii$$
Note that the coefficients in this equation coincide exactly with the orders of the zeros and the poles  of the solution of (\eqxvi) represented in the singularity pattern $\{0,\infty,0,\infty,0,\infty,0\}$. This remarkable relationship -- which in fact also holds for the case of the shorter singularity pattern $\{0,\infty,0\}$, as well as for the H-V mapping and its extension (\eqxii) --  obviously merits further investigaton.

Putting $f_n=\lambda^n$ in (\eqxvii) we obtain the characteristic equation $(\lambda^2-\lambda+1)(\lambda^4-\lambda^3-\lambda^2-\lambda+1)=0$. It turns out that the largest root of this characteristic polynomial can be computed exactly. We find $\lambda={1+\sqrt{13}\over 4}+\sqrt{{\sqrt{13}-1\over 8}}$, the numerical value of which is 1.7220838$\dots$ . The fact that the characteristic equation comprises two factors greatly facilitates the computation of the algebraic entropy of the case obtained by this late confinement. Clearly, the equation $\lambda^2-\lambda+1=0$ is the characteristic equation of $f_{n+1}-f_n+f_{n-1}=0$ and $f_n$ can be expressed simply as $f_n=(-1)^n(\alpha j^n+\beta j^{2n})$, in terms of the cubic root of unity $j=e^{2i\pi\over3}$. We have computed numerically the algebraic entropy for the case of late confinement with this specific choice of $f_n$ using Halburd's Diophantine approximation method [\refdef\halburd]. After 20 iterations of the mapping we found a value of $\exp(\varepsilon)\approx1.7221102\dots$ converging nicely towards the exact value obtained above. 
Many more examples like the present one have been worked out, all of them in agreement with the present approach.
\medskip
{\sl Conclusion}
\smallskip
Singularity confinement has been proposed as a discrete integrability criterion, based on the observation that discrete systems that are integrable by spectral methods have confined singularities. However, the discovery of the H-V example cast doubt on the usefulness of singularity confinement  as an integrability criterion. Since then, the domain of application of singularity confinement has been restricted essentially to that of deautonomisation where, starting from an integrable autonomous system, one relies on it to derive non-autonomous extensions of the original mapping.

It is precisely this deautonomisation approach that offers the possibility to make singularity confinement a reliable and efficient integrability detector. The procedure, which we have dubbed full deautonomisation, can be summarised as follows. Our starting point is a mapping with confined singularities. In order to assess its integrability we extend the system by replacing the constant coefficients in it by functions of the independent variable, which are  to be determined. The important step here is that this deautonomisation must be performed even for coefficients which are zero, i.e. for terms which are absent from the initial autonomous system but which, when included, do not modify the singularity pattern. We perform the singularity analysis on this deautonomised system requiring that the singularity pattern be exactly the same as for the initial one, obtaining thus a set of confinement constraints which fix the precise dependence of the parameters on the independent variable. If the system is integrable, the characteristic equations obtained from these constraints have only roots with modulus 1. However, if one of the characteristic equations has a root with modulus greater than 1, then this implies non-integrability. Moreover, the logarithm of the modulus of the largest root is precisely the algebraic entropy of the system.

The full deautonomisation procedure described above makes singularity  confinement  a sufficient integrability criterion. The question as to its necessary character depends on the precise definition of integrability one uses. While systems integrable by spectral methods do satisfy this criterion, there exists a class of systems, integrable through linearisation, which have unconfined singularities and to which the present discussion does not apply. We hope to be able to address these questions in more detail in the near future, and in particular the matter of proving the conjectured relation between the behaviour of the parameters in the mapping and the value of the algebraic entropy we put forward here.
\bigskip
{\scap References}
\medskip
\item{[\sincon]} B. Grammaticos, A. Ramani and V. Papageorgiou, Phys. Rev. Lett. 67 (1991) 1825.
\item{[\desoto]} B. Grammaticos, F.W. Nijhoff and  A. Ramani, {\it Discrete Painlev\'e equations}, in The Painlev\'e property -- One Century later, R. Conte (Ed.), New York: Springer-Verlag, (1999) p. 413.
\item{[\hiv]} J. Hietarinta and C-M. Viallet, Phys. Rev. Lett. 81, (1998) 325.
\item{[\mase]} T. Mase, R. Willox, B. Grammaticos and   A. Ramani, {\sl Deautonomisation by singularity confinement: an algebro-geometric justification}, preprint (2014): arXiv:1412.3883 [nlin.SI].
\item{[\take]} T. Takenawa, J. Phys. A 34 (2001) 10533.
\item{[\mimura]} N. Mimura, S. Isojima, M. Murata, J. Satsuma, A. Ramani and  B. Grammaticos, J. Math. Phys. 53 (2012) 023510.
\item{[\nevan]} M.J. Ablowitz, R.G. Halburd  and  B. Herbst, Nonlinearity 13 (2000) 889.
\item{[\tsuda]} T. Tsuda, A. Ramani, B. Grammaticos and  T. Takenawa, Lett. Math. Phys. 82 (2007) 39.
\item{[\kanki]} M. Kanki, T. Mase, T. Tokihiro, unpublished results.
\item{[\late]} J. Hietarinta and  C-M. Viallet, Chaos, Solitons and Fractals 11 (2000) 29.
\item{[\halburd]} R.G. Halburd, J. Phys. A 38 (2005) L263.
\end

\end